\documentstyle{article}

\newcommand{\ba}{\begin{eqnarray}}
\newcommand{\ea}{\end{eqnarray}}
\newcommand{\ii}{\'\i}
\begin{document}
\pagestyle{plain}

\title{A Symmetry Adapted Approach to Molecular Spectroscopy: The
Anharmonic  Oscillator Symmetry Model} 
\author{A. Frank$^{1,2)}$, R. Lemus$^{1)}$, R. Bijker$^{1)}$, 
F. P\'erez-Bernal$^{3)}$ and J.M. Arias$^{3)}$\\
\and
\begin{tabular}{rl}
$^{1)}$ & Instituto de Ciencias Nucleares, U.N.A.M.,\\
        & A.P. 70-543, 04510 M\'exico D.F., M\'exico\\
$^{2)}$ & Instituto de F\'{\i}sica, Laboratorio de Cuernavaca,\\
        & A.P. 139-B, Cuernavaca, Morelos, M\'exico\\
$^{3)}$ & Departamento de F\'{\i}sica At\'omica, Molecular y Nuclear,\\
        & Facultad de F\'{\i}sica, Universidad de Sevilla,\\
        & Apdo. 1065, 41080 Sevilla, Espa\~na
\end{tabular}}
\date{}
\maketitle
\noindent
\vspace{6pt}
\begin{abstract}
We apply the Anharmonic Oscillator Symmetry Model  to the description
of  vibrational excitations in ${\cal D}_{3h}$ and ${\cal T}_d$
molecules.  
  A systematic  procedure  can be used 
 to establish the relation between the algebraic and
configuration space formulations,  by means of which 
 new interactions are found in the
algebraic model, leading to reliable
spectroscopic predictions.  We illustrate the  method for the case of
${\cal D}_{3h}$-triatomic molecules and the ${\cal T}_d$ Be-cluster.  
\end{abstract}

\newpage

The study of molecular vibrational spectra \cite{uno} requires
theoretical models in order to analyze and interpret the
measurements \cite{dos}.  These models range from simple
parametrizations of the energy levels, such as the Dunham
expansion \cite{dos}, to {\it ab initio} calculations, where solutions
of the Schr\"odinger equation in different approximations are
sought \cite{tres,cuatro,cinco}.  In general, the latter involve the
use of internal coordinates and the evaluation of force field
constants associated to derivatives at the potential minima.  While
this method can be reliably applied to small molecules \cite{seis}, it
quickly becomes a formidable problem in the case of larger
molecules, due to the size of their configuration spaces.  New
calculational tools to describe complex molecules are thus needed.

  In
1981 an  algebraic approach was 
proposed to describe the roto-vibrational structure of diatomic
molecules \cite{siete},   subsequently  extended to  linear
tri- and 
four- atomic molecules \cite{ocho} and certain  non-linear  triatomic
molecules \cite{nueve}.  Although these   were encouraging results, the
model  could not   
be extended to  polyatomic molecules,  due to the impossibility of 
 incorporating  the  underlying 
discrete symmetries.  This difficulty could be surmounted by treating
the vibrational degrees of freedom separately from the rotations.  In
1984 Van Roosmalen {\it et al} proposed  a U(2)-based model
to describe the stretching vibrational modes in ABA
molecules \cite{diez} later 
extended to describe the stretching vibrations of polyatomic
molecules such as octahedral and benzene like molecules \cite{once}. 
Recently the bending modes have also been included in  the
framework, which was subsequently  applied to 
 describe ${\cal
C}_{2v}$-triatomic molecules \cite{doce} and the lower excitations of
tetrahedral molecules \cite{trece}, using  
a scheme which combines Lie-algebraic and point group methods.  
 In a different approach, it has also been 
 suggested  to use a $U(k+1)$ model for the $k =
3n-3$ rotational and vibrational degrees of freedom of a $n$-atomic
molecule.  This model has the advantage that it  incorporates
all rotations and vibrations and takes into  account
 the relevant point group symmetry \cite{catorce},  but  for larger
molecules the  number of possible interactions and the 
size of the Hamiltonian matrices increase very
rapidly, making  it impractical to apply. 

 Although
 the  algebraic formulations have  proved useful,  several
problems remained, most important of which is the 
absence of a clear connection to traditional methods.  On the other hand, a
related problem  is the lack of a systematic
procedure to construct all physically meaningful interactions in the
algebraic space.  In this paper we show that both these issues can be
resolved  by means of 
 a general model for the analysis of molecular vibrational
spectra,  the Anharmonic Oscillator Symmetry Model
(AOSM).  In this approach  it is possible to construct algebraic
operators with  well defined physical meaning,   in
particular  interactions  fundamental 
 for the description of the degenerate modes
present in systems exhibiting high degree of symmetry.  The
 procedure to construct them   takes full advantage of
the discrete  symmetry of the molecule and  gives rise to  all
possible terms in a 
systematic fashion,  providing  a clear-cut connection   
 between the algebraic scheme and the traditional
analyses based on internal coordinates, which  correspond to   the
harmonic limit of the model. 

As a test for this approach 
  we apply the AOSM  to the Be$_4$ cluster and to three 
${\cal D}_{3h}$-triatomic molecular systems, namely H$^+_3$, Be$_3$ and
Na$^+_3$.   Since  small molecules can in general 
be      well described by means of {\it ab initio}
calculations \cite{quince,dieciseis}, we  emphasize the basic purpose 
 of this work.
We  have established   an exact  correspondence
between configuration space and algebraic interactions by studying
the harmonic limit of the U(2) algebra.  
  This general procedure not only allows to  derive the
interactions in the AOSM  from  interactions in
configuration space, but can also  be applied to cases for which no
configuration space interactions are available.  
  The ${\cal D}_{3h}$-triatomic molecules constitute the
simplest systems where degenerate modes appear and where the new
interactions in the  model become significant. In the case of Be$_4$,
a direct comparison with {\it ab initio} calculations will be
presented.  The application
of these techniques to more complex systems, such as the methane  
molecules, is presently under investigation \cite{diecisiete}.

The   model  is based on  the isomorphism of the $U(2)$ Lie
algebra and the one dimensional Morse oscillator 
\ba
{\cal H} = - {\hbar^2 \over 2\mu} {d^2\over dx^2} + D (e^{-{2x\over
d}} - 2e^{- {x\over d}} ) ~~ ,  
\ea
whose eigenstates ${\cal E}$ can be associated with  $U(2)\supset
 SO(2)$ states \cite{dieciocho}.  In order to see how this isomorphism comes
about, consider the radial equation 
\ba
{1\over 2} \biggl( - {1\over r} {d\over dr} r {d\over dr} + {m^2
\over r^2} + r^2 \biggr) \phi (r) = (N+1) \phi (r) ~~ , 
\ea
which corresponds to a two-dimensional harmonic oscillator (in units where
$\hbar = \mu = e = 1)$ associated to a $U(2)$ symmetry
algebra \cite{diecinueve}  By carrying out the transformation 
$$
r^2 = (N+1)e^{-\rho} ~~ , 
$$
equation (2) transforms into 
\ba
\biggl[ - {d^2 \over d\rho^2} + \biggl( {N + 1 \over 2}\biggr)^2
(e^{-2\rho} - 2e^{-\rho}) \biggr] \phi(\rho) = - m^2 \phi (\rho) ~~ ,
\ea
which can be identified with (1) after defining $x = \rho d$
and  multiplying by $\hbar^2/2\mu d^2$, provided that 
\ba 
D  & = & { \hbar^2 \over 8\mu d^2} (N+1)^2 ~~ ,   \\
{\cal E} & = & - {\hbar^2 \over 2\mu d^2} m^2 ~~ . 
\ea
In the framework of the $U(2)$ algebra, the operator  $\hat N$
corresponds to the total 
number of bosons and is fixed by the potential shape according to
(4), while $m$, the eigenvalue of the $SO(2)$ generator $J_z$,
takes the values $m = \pm N/2$, $\pm (N-2)/2, \dots$.  The Morse
spectrum is reproduced twice and consequently for these applications
the $m$-values must be restricted to be positive.  In terms of the
  $U(2)$ algebra, it is  clear from (3-5) that the Morse
Hamiltonian has the algebraic realization 
\ba
\hat  H = - {\hbar^2 \over 2\mu d^2} \hat J^2_z = - A \hat J^2_z ~~.
\ea
In addition, the $U(2)$ algebra includes the raising and lowering
operators $\hat J_+$, $\hat J_-$, which connect different energy
states in (3), while the angular momentum operator is given by
$\hat J^2 = {1\over 4} \hat N ( \hat N+2)$, as can be readily shown.

The Morse Hamiltonian (6) can be rewritten in the more convenient
form  
\ba
\hat H = A \hat  H^M = {A \over 2} [ ( \hat J_ + \hat J_- + \hat J_-
\hat J_+) - \hat N ] ~~ ,  
\ea
where we have used the relation $\hat J^2_z = \hat J^2 - {1\over 2} (
\hat J_+ \hat J_- + \hat J_- \hat J_+)$ and  added  the constant
term ${A \hat N^2 \over 4}$ in order to place the ground state
at zero energy.  The parameters $N$ and $A$ appearing in (7) are
related with the usual  
 harmonic and anharmonic constants $\omega_e$ and
$x_e\omega_e$ used in spectroscopy \cite{siete}. To obtain this
relation it is convenient to introduce the quantum number 
\ba
v = {N \over 2} - m ~~ , 
\ea
which corresponds to the number of quanta in the
oscillator \cite{diecinueve}.  
This is seen by  substituting (8) into (7).  
  In terms of $v$, the corresponding energy
expression  takes the form 
\ba
E_M = -{A\over 2}(N+1/2) + A (N+1) (v + 1/2)  -A (v+1/2)^2 ~~ ,  
\ea
from which we immediately obtain 
\ba
\omega_e & = &  A (N+1) ~~ ,   \nonumber \\
x_e \omega_e & = &  A ~~ . 
\ea
Thus, in a diatomic molecule the parameters $A$ and $N$ can be 
determined by the spectroscopic constants $\omega_e$ and
$x_e\omega_e$.

We now  consider the \ $U_i(2) \ \supset \ SU_i(2) \ \supset \ SO_i(2)$ 
\ algebra, \ generated \ by \ the \ set \  
$\{ \hat G_i \} \equiv $  $ \{ \hat N_i, \, \hat J_{+,i}, 
\, \hat J_{-,i}, \, \hat J_{0,i} \}$, satisfying the commutation 
relations 
\ba
\, [ \hat J_{0,i}, \hat J_{\pm,i}] \;=\; \pm \hat J_{\pm,i} ~,
\hspace{1cm} 
\, [ \hat J_{+,i}, \hat J_{-,i}] \;=\; 2 \hat J_{0,i} ~,
\hspace{1cm} 
\, [ \hat N_i, \hat J_{\mu,i}] \;=\; 0 ~, \label{jmui}
\ea
with $\mu=\pm,0$.  As mentioned before, 
 for the symmetric irreducible representation
$[N_i,0]$  
of $U_i(2)$ one can show that the Casimir operator is given by
 \cite{diecinueve} 
$\vec{J}_i^{\, 2} = \hat N_i(\hat N_i+2)/4$, 
from which follows the identification $j_i=N_i/2$. The $SO_i(2)$
label is denoted by $m_i$. 

In the algebraic approach each relevant interatomic interaction
is associated with a $U_i(2)$ algebra \cite{once}. As a specific
example, we  
consider the Be$_4$ cluster, which has a tetrahedral shape.  ${\cal
D}_{3h}$ molecules can be similarly treated.  
 In the Be$_4$  
case there are six $U_i(2)$ algebras involved ($i=1,\ldots,6$).
 The operators in the model are expressed in terms of the 
generators of these algebras, and the symmetry requirements of the 
tetrahedral group ${\cal T}_d$  can be readily imposed
 \cite{trece,veinte}. 
The local operators $\{ \hat G_i \}$
acting on bond $i$ can be projected to any of the fundamental 
irreps $\Gamma=A_1$, $E$ and  $F_2$.
Using the $\hat J_{\mu,i}$ generators (\ref{jmui})
we obtain the ${\cal T}_d$ tensors
\ba
\hat T^{\Gamma}_{\mu,\gamma} &=& 
\sum_{i=1}^{6} \, \alpha^{\Gamma}_{\gamma,i} \, \hat J_{\mu,i} ~,
\ea
where $\mu=\pm,0$ and $\gamma$ denotes the component of $\Gamma$. 
The explicit expressions  are given by
\ba
\hat T^{A_1}_{\mu,1} &=& \frac{1}{\sqrt{6}}  
\sum_{i=1}^{6} \, \hat J_{\mu,i} ~, 
\nonumber\\ 
\hat T^{E}_{\mu,1} &=& \frac{1}{2\sqrt{3}} \left( \hat J_{\mu,1} 
+ \hat J_{\mu,2} - 2 \hat J_{\mu,3} + \hat J_{\mu,4} 
- 2 \hat J_{\mu,5} + \hat J_{\mu,6} \right) ~, 
\nonumber\\ 
\hat T^{E}_{\mu,2} &=& \frac{1}{2} \left( \hat J_{\mu,1} 
- \hat J_{\mu,2} - \hat J_{\mu,4} + \hat J_{\mu,6} \right) ~, 
\nonumber\\
\hat T^{F_2}_{\mu,1} &=& \frac{1}{\sqrt{2}} 
\left( \hat J_{\mu,1} - \hat J_{\mu,6} \right) ~,
\nonumber\\
\hat T^{F_2}_{\mu,2} &=& \frac{1}{\sqrt{2}} 
\left( \hat J_{\mu,2} - \hat J_{\mu,4} \right) ~,
\nonumber\\
\hat T^{F_2}_{\mu,3} &=& \frac{1}{\sqrt{2}} 
\left( \hat J_{\mu,3} - \hat J_{\mu,5} \right) ~. \label{tdgen}
\ea
The Hamiltonian operator can be constructed by repeated couplings 
of these tensors to a total symmetry $A_1$, since it must commute 
with all operations in ${\cal T}_d$. This is accomplished by means
of the ${\cal T}_d$-Clebsch-Gordan coefficients
 \cite{trece,veinte,veintiuno}. 

All calculations can be carried out in a symmetry-adapted basis, which 
is projected from the local basis
\ba
\begin{array}{ccccccccccccc}
U_1(2) &\otimes& \cdots &\otimes& U_6(2) &\supset& 
SO_1(2) &\otimes& \cdots &\otimes& SO_6(2) &\supset& SO(2) \\
\downarrow && && \downarrow && \downarrow && && 
\downarrow && \downarrow \\ 
| \;\; [N_1] &,& \ldots &,& [N_6] &;& 
v_1 &,& \ldots &,& v_6 &;& \; V \;\; \rangle 
\end{array}
\ea
in which each anharmonic oscillator is well defined. By 
symmetry  
considerations, $N_i=N$ for the six oscillators, $v_i= N_i/2 - m_i$
denotes  
the phonon number in bond $i$ and $V=\sum_i v_i$ is the total 
number of phonons \cite{trece,diecinueve}.
 The one-phonon states $V=1$ are denoted by 
$| \, i \, \rangle$ with $v_i=1$ and $v_{j \neq i}=0$. Using the same
projection technique as for the generators (\ref{tdgen}), we 
find the six fundamental modes
\ba
^{1}\phi^{\Gamma}_{\gamma} &=& 
\sum_{i=1}^{6} \, \alpha^{\Gamma}_{\gamma,i} \, | \, i \, \rangle ~. 
\label{tdbas}
\ea
The expansion coefficients are the same as in (\ref{tdgen}).
The higher phonon states $^{V}\phi^{\Gamma}_{\gamma}$ can also  be 
  constructed using the Clebsch-Gordan coefficients of 
${\cal T}_d$ \cite{trece,veinte}. Since all operators are expressed in
terms of  
powers of the $U_i(2)$ generators, their matrix elements can 
be easily  evaluated in closed form. The symmetry-adapted operators 
(\ref{tdgen}) and states (\ref{tdbas}) are the building blocks of the
model.  Note  that for more complex molecules, the method allows the
exact elimination of spurious states \cite{diecisiete}.

We now proceed to expicitly construct the  Be$_4$   Hamiltonian. 
For interactions that are at most quadratic in the 
generators the  procedure yields 
\ba
\hat H_0 &=& a_1 \, \hat{\cal H}_{A_1} + a_2 \, \hat{\cal H}_{E} 
+ a_3 \, \hat{\cal H}_{F_2} 
+ b_2 \, \hat{\cal V}_{E} + b_3 \, \hat{\cal V}_{F_2} ~, \label{H0}
\ea
with
\ba
\hat{\cal H}_{\Gamma} &=& \frac{1}{2N}  \left( 
\hat T^{\Gamma}_{-} \, \cdot \, \hat T^{\Gamma}_{+} +
\hat T^{\Gamma}_{+} \, \cdot \,  \hat T^{\Gamma}_{-} \right) 
\nonumber\\
\hat{\cal V}_{\Gamma} &=& \frac{1}{N}  
\hat T^{\Gamma}_{0} \, \cdot \, \hat T^{\Gamma}_{0} ~,
\ea
Note that we have not included  
$\hat{\cal V}_{A_1}$ in (\ref{H0}), since the combination 
\ba
\sum_{\Gamma} \left( \hat{\cal H}_{\Gamma} + \hat{\cal V}_{\Gamma} \right) 
&=& \frac{3}{2} (N+2) ~, 
\ea
is a constant. The five interaction terms in Eq.~(\ref{H0}) correspond 
to  linear combinations of 
 the ones obtained in lowest order in \cite{once,trece}.  
However, it is necessary  to include interactions  which are related 
to the vibrational angular momenta associated with the 
degenerate modes $E$ and $F_2$. These kind of terms is absent in the 
former versions of the model \cite{once,trece}.
We now proceed to show how they can be obtained in the AOSM.  
In configuration space the vibrational angular momentum operator 
for the $E$ mode is given by \cite{veintidos}
\ba
\hat l^{A_2} &=& -i \left( q^E_1 \, \frac{\partial}{\partial q^E_2}
- q^E_2 \, \frac{\partial}{\partial q^E_1} \right) ~,
\ea
where $q^E_1$ and $q^E_2$ are the  normal coordinates
associated to the  $E$ mode.   
This relation can be transformed to the algebraic space by means of 
the harmonic oscillator operators
\ba
b^{\Gamma \, \dagger}_{\gamma} \;=\; \frac{1}{\sqrt{2}} \left( 
q^{\Gamma}_{\gamma} - \frac{\partial}{\partial q^{\Gamma}_{\gamma}}
\right) ~, \hspace{1cm}
b^{\Gamma}_{\gamma} \;=\; \frac{1}{\sqrt{2}} \left( 
q^{\Gamma}_{\gamma} + \frac{\partial}{\partial q^{\Gamma}_{\gamma}}
\right) ~,
\ea
to obtain
\ba
\hat l^{A_2} &=& -i \left( b^{E \, \dagger}_1 b^{E}_2 - 
b^{E \, \dagger}_2 b^{E}_1 \right) ~.
\label{vibang}
\ea
Here $b^{E}_{\gamma} = \sum_{i} \alpha^E_{\gamma,i} \, b_i$, with a
similar form for $b^{\Gamma \, \dagger}_{\gamma}$, while the
$\alpha^E_{\gamma \, i}$ can be read from (13).   
In order to find the algebraic expression for $\hat l^{A_2}$  
 we first  introduce a scale transformation in (11) 
\ba
\bar b^{\dagger}_i \;\equiv\; \hat J_{-,i}/\sqrt{N_i} ~,
\hspace{1cm} 
\bar b_i \;\equiv\; \hat J_{+,i}/\sqrt{N_i} ~. \label{subst}
\ea
The relevant commutator  can be expressed as
\ba
[\bar b_i, \bar b_i^\dagger] \;=\; 
\frac{1}{N_i} [ \hat J_{+,i},\hat J_{-,i}] 
\;=\; \frac{1}{N_i} 2\hat J_{0,i} \;=\; 1 - \frac{2 \hat v_i}{N_i} ~,
\ea
where 
\ba
\hat v_i=\frac{\hat N_i}{2}-\hat J_{0,i} ~.
\ea
The other
two commutators in (11) are not modified by (22).  
In the harmonic limit, which is defined by $N_i \rightarrow \infty$,
 Eq. (23)  
reduces to the standard boson commutator  $[\bar b_i, \bar
b^\dagger_i]=1$.  
This limit corresponds to a contraction of $SU(2)$ to the Weyl algebra
and can be used to obtain a geometric interpretation of AOSM  
operators in terms of those in configuration space. 
In the opposite sense, Eq.~(\ref{subst}) provides a procedure to 
construct the  anharmonic representation of harmonic operators 
through the correspondence 
$b^{\dagger}_i \rightarrow \bar b^\dagger_i = 
 \hat J_{-,i}/\sqrt{N_i}$ and 
$b_i \rightarrow \bar b _i = \hat J_{+,i}/\sqrt{N_i}$.
Applying this method to the vibrational angular momentum 
(\ref{vibang}) we find 
\ba 
\hat l^{A_2} &=& -\frac{i}{N} \left( \hat J^E_{-,1} \hat J^E_{+,2} -
\hat J^E_{-,2} \hat J^E_{+,1} \right) ~.
\ea
For the vibrational angular momentum $\hat l^{F_1}_{\gamma}$ 
associated with the $F_2$ mode we find a similar 
expression. The AOSM  form of the corresponding  Hamiltonian  
interactions is 
\ba
\hat H_1 &=& c_2 \, \hat l^{A_2} \, \hat l^{A_2} 
+ c_3 \, \sum_{\gamma} \hat l^{F_1}_{\gamma} \, \hat
l^{F_1}_{\gamma} ~, 
\ea
With this method we can  obtain  
an algebraic realization of arbitrary configuration space
interactions. As a simple example, a one-dimensional harmonic
oscillator Hamiltonian $\hat H_i = 1/2 (b^\dagger_i b_i + b_i
b^\dagger_i)$, transforms into 
\ba
\frac{1}{2N} (\hat J_{-,i} \hat J_{+,i} + \hat J_{+, i} \hat J_{-,i})
= \frac{1}{N} (\hat J^2_i - \hat J^2_{0,i}) = \hat v_i + 1/2 - \frac{
\hat v^2_i}{N} ~,
\ea
where in the last step we used relation (24).  The spectrum of (27)
has an anharmonic correction, analogous to the quadratic term in the
Morse potential spectrum.  We are thus substituting harmonic
oscillators by Morse oscillators in the AOSM.

A more interesting application is to use  
 our model to fit the spectroscopic data of several polyatomic
molecules.  In the case of Be$_4$ the energy spectrum  
was analyzed by {\it ab initio} methods in \cite{quince}, where force-field
constants corresponding to an expansion of the potential up to fourth
order in the normal coordinates and momenta were evaluated. We have
generated the {\it ab initio} spectrum up to three phonons using the
analysis in   
 \cite{veintidos}. For the algebraic Hamiltonian we take
\ba
\hat H &=& \hat H_0 +  
c_3 \, \sum_{\gamma} \hat l^{F_1}_{\gamma} \, \hat l^{F_1}_{\gamma} 
+ d_3 \, (\hat{\cal H}_{F_2}+\hat{\cal V}_{F_2})^2 
+ e_3 \, \hat{\cal O}_{33} ~, 
\ea
where $\hat H_0$ was defined  in (16).  
The  term $\hat{\cal O}_{33}$ is the algebraic form of the 
corresponding $F_2$-mode  interaction in \cite{veintidos} which 
is responsible for the splitting of   the vibrational levels with the
same  $l$ in the $F_2$ overtones.  For ${\cal D}_{3h}$ molecules we
can follow an analogous procedure, namely, we can construct the
${\cal D}_{3h}$ symmetry-adapted operators and states corresponding
to (13) and (15) and carry out the building up procedure to construct
the Hamiltonian and higher phonon states, using in this case the
appropriate projection operators and Clebsch-Gordan
coefficients \cite{trece,veinte}.  Here we omit the details for lack
of space and only present the fit to the energy
spectrum \cite{veintitres,veinticuatro}. 

Note that the Be$_4$ Hamiltonian (28) preserves the total
phonon-number $V$.  This is a good approximation for this case
according to the analysis of \cite{veintidos,veinticinco}, but it is
known that Fermi  
resonances can occur for certain molecules when the fundamental mode
frequencies are such that $(V,V^\prime)$ states with $V \neq
V^\prime$ are close in energy.  These interactions can be introduced
in the Hamiltonian but the size of the energy matrices grows very
rapidly, so the best way to deal with this problem is through
perturbation theory.  We now present the results of our least-square
fits to the energy spectra of Be$_4$, Be$_3$, Na$^+_3$ and H$^+_3$.

In Table I we show the fit to Be$_4$ using the Hamiltonian (28).  The
r.m.s. deviation obtained is 2.3 cm$^{-1}$, which can be considered
of spectroscopic quality.  We point out that in
 \cite{veintidos,veinticinco}  several
higher order interactions are present which we have  neglected.
Since our model  can be put into a one to one correspondence with the
configuration space calculations, it is in fact possible to improve
the accuracy of the fit considerably, but we have lused a simpler
Hamiltonian than the one of \cite{veintidos,veinticinco}.  When no
{\it ab initio} 
calculations are available (or feasible) the AOSM approach can be
used empirically, achieving increasingly good fits by the inclusion
of higher order interactions \cite{diecisiete}.  In Table II we
presenlt AOSM fits to the spectra of Be$_3$, Na$^+_3$ and H$^+_3$ up
to three phonons.  While remarkably accurate descriptions of the
first two molecules can be achieved using a four-parameter
Hamiltonian, we were forced to include four additional higher order
terms in the H$^+_3$ Hamiltonian in order to properly describe this
molecule.  This is in accordance with the work of Carter and
Meyer \cite{dieciseis}, who were forced to include twice as many
terms in the potential energy surface lfor H$^+_3$ than for the
Na$^+_3$ molecule.  The H$^+_3$ ion is a very ``soft'' molecule
which, due to the light mass of its atomic constituents carries out
large amplitude oscillations from its equilibrium
positions \cite{dieciseis}.

  A finer test for 
the model is to use the  
wave functions to evaluate  infrared and Raman transitions. 
The algebraic realization of the transition operators can be obtained
from 
their  expression in configuration space using the large $N$
connection,  or purely algebraically by  their  
tensorial properties under the relevant point group \cite{veinticuatro}. 
 We remark  that the model can also be extended to include the 
rotational degrees of freedom, by coupling the vibrational wave 
functions to rotational states properly symmetrized to carry the 
point group representations \cite{veintidos}.

 The AOSM  is based on symmetry
methods which systematically incorporate group theoretical
techniques, providing a clear methodological procedure that can be
applied to more complex molecules.  We define symmetry adapted
operators that have a
specific action 
over the function space.  This is a general procedure which gives
rise to a clear physical interpretation of the interactions and has
the additional advantage of 
considerably improving the convergence during the least square energy
fits.  Based on the harmonic limit of the SU(2) algebra
we have found a systematic approach to derive an algebraic
realization of interactions given in configuration space.  The model 
 surmounts one of the main objections raised against the use of
algebraic
models, where it was not possible to obtain a direct correspondence
with the configuration-space 
  approaches.  
  For the general case  when there is no
information about the 
 form  of these interactions in configuration space, we have
devised an  algebraic procedure to derive them  using their 
tensorial  structure  under the point group.  The
combination of the different methodologies leads to the AOSM, 
 which can be  applied in the same fashion to more complex molecules.

We believe that the  AOSM  represents a systematic,
simple but accurate alternative to the traditional methods,
particularly for polyatomic molecules, where the integro-differential
approaches are too complex to be applied or require very large 
numerical calculations.  Since the model provides manageable wave
functions, it is possible to evaluate the matrix elements of
arbitrary physical operators, which have a simple representation in
the algebraic space.   The study of electromagnetic intensities, as
well as the application of the model to more complex molecules is
currently under investigation  \cite{diecisiete,veinticuatro}.

\

This work was supported in part by the 
European Community under contract nr. CI1$^{\ast}$-CT94-0072, 
DGAPA-UNAM under project IN105194, CONACyT-M\'exico under project 
400340-5-3401E and Spanish DGCYT under project PB92-0663.

\begin{table}
\centering
\caption[]{\small 
Fit to ab initio \cite{once} calculations for Be$_4$. The values of the 
parameters are $N = 40$, $a_1=646.95$, $a_2=463.51$, $a_3=675.54$,  
$b_2=129.46$, $b_3=317.01$, $c_3=-14.285$, $d_3=10.688$, and
$e_3=-1.948$.  
The parameters and energies are given in cm$^{-1}$.
$(\nu_1,\nu_2^m,\nu_3^l)$ denotes the number of phonons 
in the $A_1$, $E$ and $F_2$, respectively, and $m$ and $l$ 
the value of the vibrational angular momenta \cite{dieciseis}.
\normalsize}
\vspace{10pt} \label{fit} 
\begin{tabular}{cclrr|cclrr}
\hline
& & & & & & & & & \\
$V$ & $(\nu_1,\nu_2^m,\nu_3^l)$ & $\Gamma$ 
& Ab initio & Present &
$V$ & $(\nu_1,\nu_2^m,\nu_3^l)$ & $\Gamma$ 
& Ab initio & Present \\
& & & & & & & & & \\
\hline
& & & & \\
1 & $(1,0^0,0^0)$ & $A_1$ &  638.6 &  638.3 & 
3 & $(1,0^0,2^0)$ & $A_1$ & 2106.8 & 2110.1 \\
  & $(0,1^1,0^0)$ & $E$   &  453.6 &  454.8 & 
  & $(1,0^0,2^2)$ & $E$   & 2000.1 & 2005.2 \\
  & $(0,0^0,1^1)$ & $F_2$ &  681.9 &  681.1 & 
  &               & $F_2$ & 2056.8 & 2058.6 \\
2 & $(2,0^0,0^0)$ & $A_1$ & 1271.0 & 1271.2 & 
  & $(0,3^1,0^0)$ & $E$   & 1341.3 & 1347.1 \\
  & $(1,1^1,0^0)$ & $E$   & 1087.1 & 1085.1 & 
  & $(0,3^3,0^0)$ & $A_1$ & 1355.5 & 1352.3 \\
  & $(1,0^0,1^1)$ & $F_2$ & 1312.6 & 1312.8 & 
  &               & $A_2$ & 1355.5 & 1353.0 \\
  & $(0,2^0,0^0)$ & $A_1$ &  898.3 &  903.0 & 
  & $(0,2^{0,2},1^1)$ & $F_2$ & 1565.5 & 1565.4 \\
  & $(0,2^2,0^0)$ & $E$   &  905.4 &  905.7 & 
  &                   & $F_2$ & 1584.4 & 1583.3 \\
  & $(0,1^1,1^1)$ & $F_2$ & 1126.7 & 1126.1 & 
  & $(0,2^2,1^1)$ & $F_1$ & 1578.5 & 1576.2 \\
  &               & $F_1$ & 1135.5 & 1134.8 & 
  & $(0,1^1,2^{0,2})$ & $E$   & 1821.4 & 1821.9 \\
  & $(0,0^0,2^0)$ & $A_1$ & 1484.0 & 1485.8 & 
  &                   & $E$   & 1929.5 & 1928.6 \\
  & $(0,0^0,2^2)$ & $E$   & 1377.3 & 1378.1 & 
  & $(0,1^1,2^2)$ & $A_1$ & 1813.3 & 1814.8 \\
  &               & $F_2$ & 1434.1 & 1433.6 & 
  &               & $A_2$ & 1830.8 & 1830.7 \\
3 & $(3,0^0,0^0)$ & $A_1$ & 1897.0 & 1898.8 & 
  &               & $F_1$ & 1874.5 & 1873.9 \\
  & $(2,1^1,0^0)$ & $E$   & 1714.3 & 1710.0 & 
  &               & $F_2$ & 1883.2 & 1881.9 \\
  & $(2,0^0,1^1)$ & $F_2$ & 1937.0 & 1939.1 & 
  & $(0,0^0,3^{1,3})$ & $F_2$ & 2136.5 & 2137.5 \\
  & $(1,2^0,0^0)$ & $A_1$ & 1526.6 & 1525.5 & 
  &                   & $F_2$ & 2327.3 & 2328.1 \\
  & $(1,2^2,0^0)$ & $E$   & 1533.7 & 1527.8 & 
  & $(0,0^0,3^3)$ & $F_1$ & 2199.8 & 2200.2 \\
  & $(1,1^1,1^1)$ & $F_2$ & 1752.2 & 1750.1 & 
  &               & $A_1$ & 2256.5 & 2257.5 \\
  &               & $F_1$ & 1761.0 & 1758.1 & 
  &               &       &        & \\
& & & & & & & & & \\
\hline
\end{tabular}
\end{table}

\begin{table}
\centering
\caption[]{\small Least square energy fit for H$^+_3$, Be$_3$
and Na$^+_3$  using the  Hamiltonian (7.1).  We show
 the energy differences $\Delta E = E_{th} - E_{exp}$.  The values of
the energies   $E_{exp}$
are given in Table V. 
\normalsize}
\vspace{10pt} 
\begin{tabular}{|ccrrr|}
\hline
 & & & & \\
& & H$^+_3$ & Be$_3$ & Na$^+_3$ \\
$(vA_1 v^l_E)$ & Symmetry & $\Delta E$ & $\Delta E$ & $\Delta E$ \\ 
& & & &  \\
$(01^1)$ & $e$ & -1.55 &  0.51 &  0.93 \\ 
$(10^0)$ & $a_1$ &  0.42 &  0.02 &  1.95 \\ 
& & & &  \\
$(02^0)$ & $a_1$ & 7.48 &  -0.74 &  0.37 \\ 
$(02^2)$ & $e$ & -5.69 &  0.17 &  0.84 \\
$(11^1)$ & $e$ & -0.61 &  0.82 &  1.68 \\
$(20^0)$ & $a_1$ & -0.11 &  -0.04 &  1.26 \\
& & & &  \\ 
$(03^1)$ & $e$ & -4.46 &  - 2.05 &  -1.19 \\ 
$(03^3)$ & $a_1$ & 3.18 &  -1.23 &  -0.34 \\
$(03^3)$ & $a_2$ & 2.44 &  0.61 &  -0.33 \\
$(12^0)$ & $a_1$ & 0.66 &  1.90 &  -0.01 \\
$(12^2)$ & $e$ & -5.0 &  -1.36 &  0.34 \\
$(21^1)$ & $e$ & 4.07 &  0.79 &  -0.19 \\
$(30^0)$ & $a_1$ &  -1.23 &  -1.66 &  -2.06 \\
& & & &  \\
\hline 
& & & &  \\ 
\multicolumn{2}{|r} \, rms & 3.6  & 0.98 &   1.10 \\
& & & &  \\
\multicolumn{2}{|r} \, Number of Parameters & 8 & 4 & 4 \\
& & & &  \\
\hline
\end{tabular}
\end{table}

\end{document}